\documentclass[
    ,final            
  ]
  {aipproc}

\layoutstyle{6x9}

\newcommand{\beq}{\begin{equation}}
\newcommand{\eeq}{\end{equation}}
\newcommand{\bea}{\begin{eqnarray}}
\newcommand{\eea}{\end{eqnarray}}

\begin{document}

\title{NLO BFKL in $\gamma^{*} \gamma^{*}$ collisions }


\author{F. Caporale}{
  address={\it Dipartimento di Fisica, Universit\`a
della Calabria \\
and Istituto Nazionale di Fisica Nucleare, Gruppo collegato di Cosenza \\
I-87036 Arcavacata di Rende, Cosenza, Italy}
}

\author{D.Yu Ivanov}{
  address={Sobolev Institute of Mathematics and Novosibirsk State University, \\ 
630090 Novosibirsk, Russia}
}

\author{A. Papa}{
  address={\it Dipartimento di Fisica, Universit\`a
della Calabria \\
and Istituto Nazionale di Fisica Nucleare, Gruppo collegato di Cosenza \\
I-87036 Arcavacata di Rende, Cosenza, Italy}}

\begin{abstract}
 We study in the BFKL approach the total hadronic cross section for the collision of 
two virtual photons for energies in the range of LEP2 and of future linear colliders.
The BFKL resummation is done at the next-to-leading order in the BFKL Green's 
function; photon impact factors are taken instead at the leading order, but with
the inclusion of the subleading terms required by invariance under changes of the 
renormalization scale and of the BFKL scale $s_0$. We compare our results with 
previous estimations based on a similar kind of approximation.
\end{abstract}

\maketitle


\section{Introduction}

  The total hadronic cross section for the collision of two off-shell photons
with large virtualities is a fundamental observable, since it is fully under
the control of perturbative QCD.
It is widely believed that this total cross section is the 
best place for the possible manifestation of the BFKL dynamics~\cite{BFKL}
at the energies of future linear colliders (for a review, see 
Ref.~\cite{Wallon:2007xc}). For this reason, many 
papers~\cite{photons_BFKL} have considered the inclusion of the BFKL resummation
of leading energy logarithms. In a remarkable paper~\cite{Brodsky:2002ka} (see also
Ref.~\cite{Brodsky:1998kn}), 
BFKL resummation effects have been taken into account also at the subleading order
and evidence has been presented that the appearance of BFKL dynamics is compatible
with experimental data already at the energies of 
LEP2~\cite{Achard:2001kr,Abbiendi:2001tv}.
In this work~\cite{paper} we estimate the energy dependence of the $\gamma^* \gamma^*$ 
total hadronic cross section in an energy range which covers
LEP2 and future linear colliders. The procedure we follow is approximate, since
we use the singlet forward NLA BFKL Green's function together with forward 
$\gamma^* \to \gamma^*$ impact factors {\em at the leading order}. 
However, in the impact factors we include the subleading terms required by the 
invariance of the full amplitude at the NLA under change of the renormalization
scale and of the energy scale $s_0$ entering the BFKL approach.

\section{The $\gamma^* \gamma^*$ total cross section: numerical analysis}

The total hadronic cross section of two unpolarized photons with virtualities 
$Q_1^2$ and $Q_2^2$ can be obtained from the imaginary part of the forward amplitude.
Following the procedure of Refs.~\cite{mesons_1-2}, it is possible
to write down the cross section with the inclusion of NLO corrections in the Green's
function only, while keeping the impact factors at the LO. In fact, the requirement of invariance of the amplitude at the NLA under renormalization group transformation and under change of the energy scale 
$s_0$ allows to fix the $\mu_R$- and $s_0$-dependent terms in the NLO impact 
factors
\begin{equation}
A(s_0) = \chi(\nu) \ln\frac{s_0}{Q_1 Q_2}\;, 
\;\;\;\;\;\;\;\;\; 
B(\mu_R)= \frac{\beta_0}{2N_c} \ln\frac{\mu_R^2}{Q_1 Q_2}\;.
\label{AB}
\end{equation}
The details of the analytical calculation can be found in Ref.~\cite{paper}.
We use the series representation~\cite{mesons_1-2}, that is one of the 
infinitely many possible ways, equivalent with NLA accuracy, to represent the
total cross section. In the case of the 
$\gamma^*\gamma^*\to VV$ process~\cite{mesons_1-2}, where $V$ stands for light
vector meson ($\rho$, $\omega$, $\phi$), it turned out that the contribution
to the amplitude from the kernel starts to dominate over that from the impact factors in the series from
$n=4$. This makes evident the fact that the high-energy behavior of the amplitude 
is weakly affected by the NLO corrections to the impact factor. Therefore, our
approximated $\gamma^*\gamma^*$ total cross section should compare better
and better with the correct result as the energy increases.
In order to stabilize the perturbative 
series, it is necessary to resort to some optimization procedure, exploiting the 
freedom to vary the energy parameters, $\mu_R$ and $s_0$, without corrupting the 
calculation but at the next-to-NLA. We use both 
the principle of minimal sensitivity (PMS method)~\cite{Stevenson} and the 
Brodsky-Lepage-Mackenzie (BLM) method~\cite{BLM}: for some selected values of the 
energy $s$ in the region of interest the optimal scales $\mu_R$ and $s_0$ are found 
and the cross section is thus determined. Then, the curve giving the cross section 
{\it vs} the energy is obtained by interpolation.

\begin{figure}
\begin{minipage}{.48\textwidth} 
  \includegraphics[height=.2\textheight]{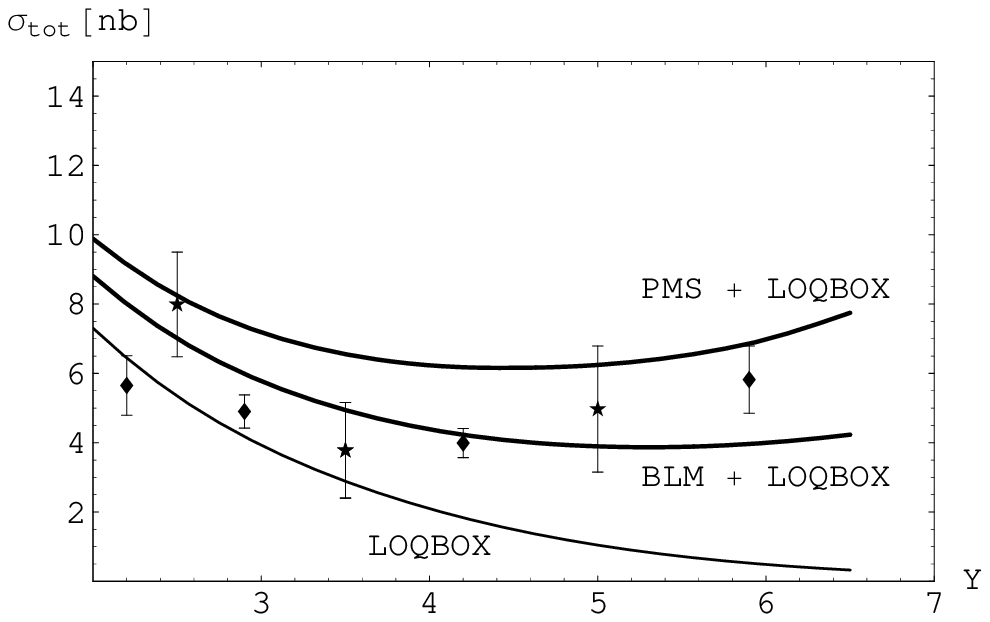}
\end{minipage}
\hspace{0.45cm}
\begin{minipage}{.48\textwidth} 
\includegraphics[height=.2\textheight]{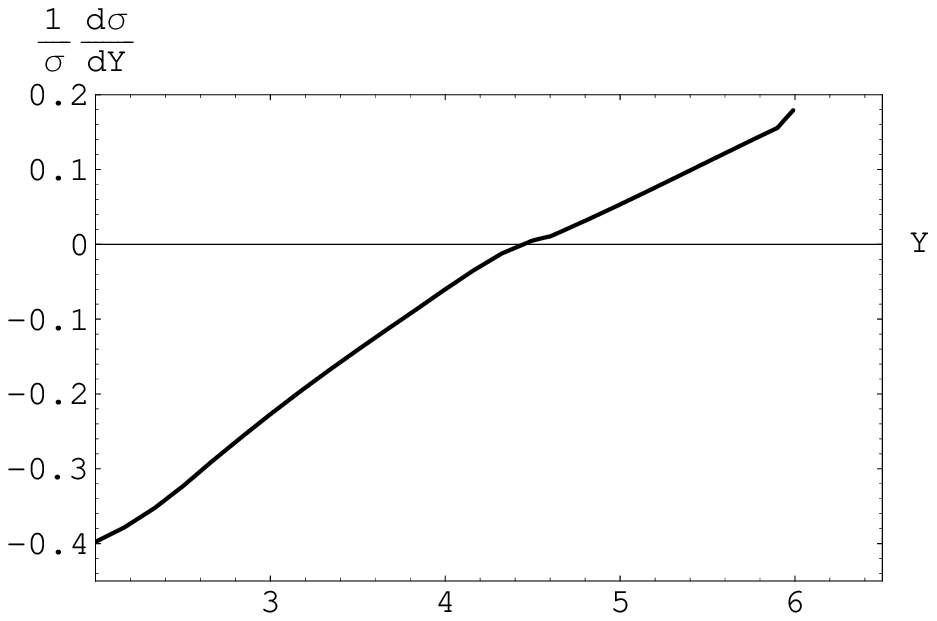}
\end{minipage}
  \caption{(Left) Energy dependence of the total cross on section on $Y\equiv\ln(s/Q^2)$.  For comparison,
experimental data from OPAL~\cite{Abbiendi:2001tv} (stars, $Q^2$=18~GeV$^2$) 
and L3~\cite{Achard:2001kr} (diamonds, $Q^2$=16~GeV$^2$) are shown. (Right) $Y$-dependence of the Pomeron intercept (minus 1) calculated from 
the total cross section with the PMS method. All theoretical curves are obtained for $Q^2$=17~GeV$^2$ and $n_f=4$.}
\label{fig:sigma}
\end{figure}

In order to compare the theoretical prediction with the existing data from LEP2, 
we cannot neglect the contribution from LO quark box diagrams~\cite{Budnev:1974de}
which is of order $\alpha^2 (\ln s)/s$. On the other hand, the soft Pomeron contribution, 
if estimated within the vector-dominance model, is proportional to 
$\sigma_{\gamma^*\gamma^*} \sim (m^2_V/Q^2)^4 \sigma_{\gamma \gamma}$ and 
is therefore suppressed for highly virtual photons.
We restrict ourselves to the case of a symmetric kinematics, which means the same 
virtuality $Q_1 =Q_2 \equiv Q$ for the two photons. This is the so-called
``pure BFKL regime'', as opposite to the ``DGLAP regime'' realized for strongly
ordered photon virtualities. In Fig.~\ref{fig:sigma} (left) we summarize our results for the CERN LEP2 region: 
we show the NLO BFKL curves obtained 
by the PMS and the BLM methods, to which we added the contribution of the LO 
quark box diagrams. For comparison we put in this plot also the experimental 
data from CERN LEP2, namely three data points from OPAL~\cite{Abbiendi:2001tv} 
($Q^2$=18~GeV$^2$) 
and four data points from L3~\cite{Achard:2001kr} ($Q^2$=16~GeV$^2$).
We observe first of all that the difference between the two theoretical
curves can be taken as an estimate of the systematics effects which underlay
the optimization procedures adopted here. Then, the comparison with experimental 
data is acceptable, although within uncertainties which are large both on the theory
and on the experiment side. Around the energy for which the condition for the
BFKL resummation, $\bar \alpha_s Y \sim 1$, is satisfied, which in the present
conditions corresponds to $Y\sim 5$, {\em both} the PMS and BLM curves agree with
experimental data within the errors. Finally, we remark that the determination from 
Ref.~\cite{Brodsky:2002ka} falls between our two curves from PMS and BLM methods.
From the energy dependence of the NLO BFKL cross section determined through the
PMS method at $Q^2$=17~GeV$^2$ we obtained also the ``dynamical'' Pomeron intercept
(minus 1) as a function of the energy. 

The result is shown in Fig.~\ref{fig:sigma} (right).
In Fig.~2 we show the $Y$-behavior of the total cross section for 
$Q^2$=20~GeV$^2$ ($n_f$=5) in an energy region not explored by past and present 
experiments, but relevant for future colliders. We plot here the two curves 
obtained in the present work with the PMS and the BLM methods. The condition 
for the BFKL resummation, $\bar \alpha_s Y\sim 1$, corresponds here to $Y\sim 6$; 
around this energy the deviation between the PMS and the BLM methods is about $50\%$. 
This discrepancy can be taken as an estimate of the systematic uncertainty of 
this approach. We observe that our determination from the BLM method is
in quite good agreement with the result of Ref.~\cite{Brodsky:2002ka} (see Fig.~4
of that paper), obtained for the same kinematics.

\section{Conclusions}

We have found that, if suitable methods are used to stabilize the perturbative
series, a smooth curve for the energy behavior of the cross section can be
achieved.
Our result in the CERN LEP2 region compares favorably with experimental
data. Systematic effects coming from the optimization procedure are estimated 
by the comparison with two different methods. Our findings in the CERN LEP2 region 
are in agreement with the result of Ref.~\cite{Brodsky:2002ka}, where for the first 
time subleading BFKL effects were considered in the $\gamma^* \gamma^*$ total 
hadronic cross section. 
In comparison with the latter work, which deals with the same process
under consideration in the present paper, the elements of novelty are the
following:
\begin{itemize}
\item the optimization procedures to stabilize the perturbative series are performed
on the amplitude itself and not on the NLO Pomeron intercept, which is not 
a physical quantity;
\item the impact factors, although taken at the LO, contain the appropriate
NLO terms, so that the dependence on the energy scales entering the process 
(the renormalization scale $\mu_R$ and the parameter $s_0$ introduced in the BFKL 
approach) is pushed to the next-to-NLA;
\item two optimization methods are used, thus having a control of systematic 
effects at work.
\end{itemize}
\begin{figure}

  \includegraphics[height=.2\textheight]{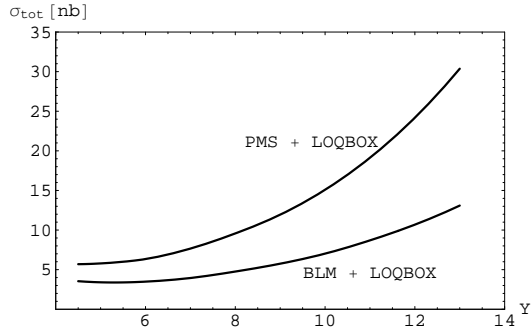}

  \caption{The $Y$-dependence of $\sigma_{\gamma^*\gamma^*}$ [nb] for virtual photon - photon collisions
at $Q^2$=20~GeV$^2$ and $n_f=5$, obtained with the PMS and the BLM methods.}
\label{sigma2}
\end{figure}

The numerical effect of the neglected subleading corrections to the impact factors 
cannot be quantified. We expected that it be modest in the second region of
energy considered in this work. Here our prediction should be very close
to the true NLO result.

The final word will be said when the $\gamma^* \gamma^*$ cross section will be 
calculated fully in the next-to-leading approximation.

\end{document}